\documentclass[12pt]{article}
\usepackage{amsmath}
\usepackage{amsfonts}
\usepackage{amssymb}
\usepackage{graphicx}
\usepackage{amsthm}
\usepackage{float}
\usepackage{subcaption}
\usepackage[iso,american]{isodate}
\usepackage[bookmarks=false]{hyperref}
\usepackage [margin=1in]{geometry}
\usepackage{enumerate}
\usepackage{listings}
\usepackage{natbib}

\hypersetup{pdfborder={0 0 0}, colorlinks=true, urlcolor=black, citecolor=black}

\newtheorem{theorem}{Theorem}[section]
\newtheorem{proposition}[theorem]{Proposition}
\newtheorem{definition}[theorem]{Definition}

\newcommand{\Dk}{D^{(x,k+1)}}

\DeclareMathOperator*{\argmin}{arg\,min}

\title{Bayesian Trend Filtering}
\date{\today}
\author{Edward A. Roualdes}

\begin{document}
\maketitle

\begin{abstract}
We develop a fully Bayesian hierarchical model for trend filtering, itself a new development in nonparametric, univariate regression.  The framework more broadly applies to the generalized lasso, but focus is on Bayesian trend filtering.  We compare two shrinkage priors, double exponential and generalized double Pareto.  A simulation study, comparing Bayesian trend filtering to the original formulation and a number of other popular methods shows our method to improve estimation error while maintaining if not improving coverage probability.  Two time series data sets demonstrate Bayesian trend filtering's robustness to possible violations of its assumptions.

\end{abstract}

\section{Introduction} 
\label{sec:intro}
Consider the nonparametric model of the function $f_0: [0,1] \mapsto \mathbb{R}$, with zero mean, independent, Gaussian errors $\epsilon_i$
\begin{equation}
  \label{eq:model}
  y_i = f_0(x_i) + \epsilon_i, \quad i = 1, \ldots, n.
\end{equation}
\noindent Assume the observations $y_i$ are generated via $f_0$ from the unique inputs $x_1 < \cdots < x_n$.  \citet{Mammen:1997} propose an estimator of $f_0$, convergent at the minimax rate, by penalizing the total variation of the $k^{th}$ derivative of the function $f$, taken to be \[ \text{TV}(f^{(k)}) = \sum_{i=1}^p |f^{(k)}(t_{i+1}) - f^{(k)}(t_i)|, \] with the set of knots $\{t_1, \ldots, t_{p+1}\}$ equal to the inputs.  Since the penalty TV$(f^{(k)})$ is not easily computed, an alternative solution uses an estimate of the total variation penalty.  Consider the estimator $\hat{f} = (\hat{f}(x_1), \ldots, \hat{f}(x_n))^t$ of $f_0 = (f_0(x_1), \ldots, f_0(x_n))^t$ that solves \[ \hat{f} = \argmin_{f}  ||y - f||_2^2 + \lambda \widehat{\text{TV}}(f^{(k)}),\] where $y$ is taken to be the vector of responses.  R.\ J.\ \citet{Tibshirani:2014} proposed such an estimator that maintains the optimal (minimax) convergence rate.  His method, called trend filtering, estimates TV$(f^{(k)})$ using the divided difference of order $k$ of $f$.  \citet{DeVore:1993} and \citet{Boor:2005} provide two excellent references on the divided difference.

Trend filtering uses the generalized lasso framework \citep{Tibshirani:2014}.  Assume $y \in \mathbb{R}^{n}$ and a model matrix $X \in \mathbb{R}^{n \times p}$.  The generalized lasso finds the minimum vector $\beta \in \mathbb{R}^{p}$ constrained by a linear transformation $D \in \mathbb{R}^{m \times p}$
\begin{equation}
  \label{eq:obj}
  \hat{\beta} = \argmin_{\beta}||y - X\beta||_2^2 + \lambda ||D\beta||_1,
\end{equation}
\noindent for some penalty parameter $\lambda > 0$.  Trend filtering uses the minimization problem in Equation~(\ref{eq:obj}) with $X:= I_n$ and by cleverly choosing a penalty matrix $D$ that when coupled with the $\ell_1$ norm recovers an estimate of the total variation penalty.

We adapt trend filtering, as a special case of the generalized lasso, to the Bayesian setting.  Similar efforts to fit lasso type, or other general shrinkage, problems with a hierarchical Bayesian model have been made \citep[see][]{Park:2008, Hans:2009, Kyung:2010, Griffin:2011, Lee:2012}.  We further investigate Bayesian trend filtering with another shrinkage prior, the generalized double Pareto, as developed by \citet{Lee:2010} and \citet{Armagan:2013}.  The generalized double Pareto prior, when applied to Bayesian trend filtering, appears to estimate the true, underlying function with smaller error and to provide better frequentist coverage properties than does the more standard double exponential prior.

This paper reads as follows. Section~\ref{sec:tf} discusses trend filtering, its minimax convergence rate and standard errors.  We present Bayesian trend filtering and details on its implementation in Section~\ref{sec:bayes}.  A simulation study comparing original trend filtering to the Bayesian version, and both of these to some popular univariate, nonparametric regression methods, is carried out in Section~\ref{sec:sims}.  Section~\ref{sec:real} fits these methods to two real data sets.  Section~\ref{sec:discussion} briefly mentions Bayesian trend filtering's relation to Gaussian process regression, summarizes our findings, and mentions some future research directions.  

\section{Trend Filtering}
\label{sec:tf}

Trend filtering, as discussed here, was developed in two stages.  \citet{Kim:2009} first introduced $\ell_1$ trend filtering for piecewise linear fits, providing a primal-dual interior point algorithm to fit the method at a specified value of the penalty parameter, $\lambda$.  \citet{Tibshirani:2011} next identified trend filtering as a special case of the generalized lasso, thus offering a path algorithm that fits trend filtering over all values of the penalty parameter, $\lambda \in [0,\infty)$.  In a separate paper \citet{Tibshirani:2014} established convergence properties of the method.   In the end, trend filtering fits a piecewise polynomial of order $k$ with knots taken to be the set of inputs $\{x_i\}_i^n$.  The piecewise polynomial comes from a continuous-time representation of the following discrete minimization problem,

\begin{equation}
  \label{eq:tf}
  \hat{f} = \argmin_{f}||y - f||_2^2 + \lambda||\Dk f||_1.
\end{equation}

\noindent The discrete difference matrix $\Dk$ depends both on the order of the derivative of the function $f$, as chosen by the practitioner, and the inputs $x_i, \, i=1, \ldots, n$.  The penalty term $||\Dk f||_1$ estimates the total variation of the $k$th derivative of the function $f$.  Because of this close relation to locally adaptive regression splines, trend filtering adapts to the fluctuations of the underlying curve and thus achieves the same optimal convergence rate \citep{Tibshirani:2014}.

\subsection{Minimax Convergence Rate}
Trend filtering converges at the minimax rate to the true underlying function of interest, as is shown by \cite{Tibshirani:2014}.  He used purely algebraic methods to show that trend filtering is sufficiently close to the locally adaptive regression spline estimator for the minimax convergence rate to carry over to trend filtering.  It is well known that this convergence rate is not achieved by any estimator linear in $y$ \citep{Donoho:1998,Tibshirani:2014}.  Thus, the optimal convergence rate depends on the $\ell_1$ norm.

Another strategy to prove the minimax convergence rate for trend filtering would make use of the metric entropy of the underlying function space for which trend filtering finds its solution.  This strategy is analogous to that of \citet{Mammen:1997} for locally adaptive regression splines.  \citet{Tibshirani:2014} briefly mentions the difficulty of such a proof.  The requisite interpolating properties of trend filtering are quite difficult to establish because trend filtering uses piecewise polynomials with potentially discontinuous lower order derivatives.  We mention this alternative strategy, not because we were able to over-come the aforementioned difficulties, but instead to highlight how Bayesian trend filtering could share metric entropy proofs of convergence rates across penalized regression and Gaussian process regression methods.  This connection is discussed further in Section~\ref{sec:discussion}.

\subsection{Standard Errors}
\label{sec:se}

\begin{figure}[H]
  \centering
  \begin{subfigure}[b]{0.48\textwidth}
    \includegraphics[width=\textwidth]{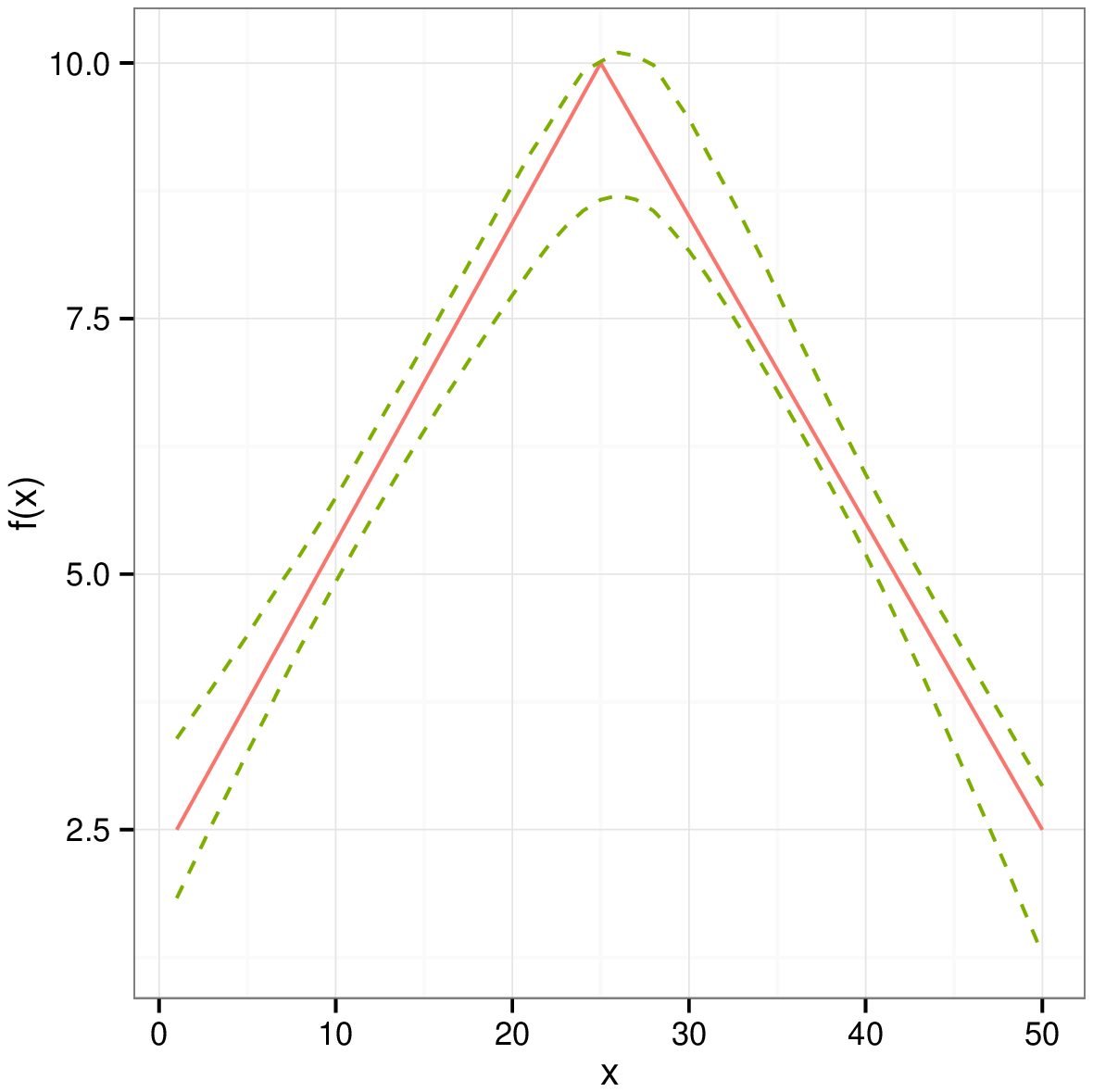}
  \end{subfigure}
  ~
  \begin{subfigure}[b]{0.48\textwidth}
    \includegraphics[width=\textwidth]{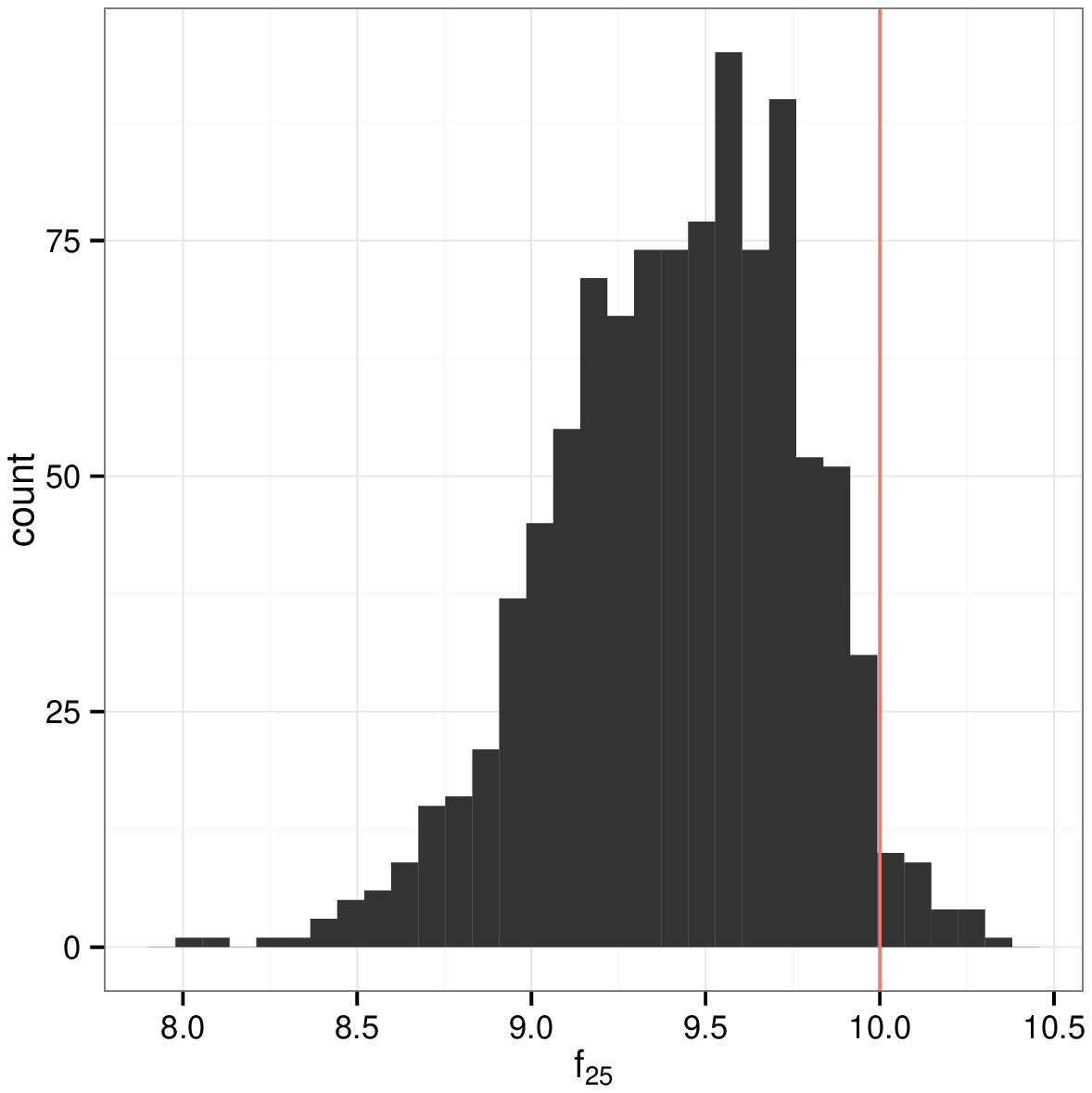}
  \end{subfigure}
  \caption{Left: A true piecewise linear function (solid red) with $95$\% bootstrap confidence intervals of a trend filtering fit, with $\hat{\lambda}_{CV(10)}$ chosen by $10$-fold cross validation.  Right: A histogram of the $500$ bootstrapped values of $\hat{f}_{25}$, with the true value of $f_{25}$ denoted by a red vertical line.}
  \label{fig:boot}
\end{figure}

As is summarized in \cite{Kyung:2010}, estimating the standard errors of lasso problems is quite difficult.  This is also seen by the significant amount of attention paid to the problem; for example, see \citep{Tibshirani:1996,Knight:2000,Osborne:2000,Fan:2001,Potscher:2009}.  However, with trend filtering the theory cited here does not find easy evidence.  Figure~\ref{fig:boot} displays $2.5$\% and $97.5$\% quantiles of bootstrapped function evaluations at each input, for a simple piecewise linear function (solid red).  The $95$\% confidence intervals (green dash) provide seemingly reasonable estimates of error.  The only obvious problem seen in Figure~\ref{fig:boot} is the bias indicated by the histogram of the bootstrapped estimates of $\hat{f}_{25}$.  

Cross validation, the recommended method to estimate $\lambda$, is both computationally costly and is well known to encourage over-fitting \citep{Davison:1997}.  If you are willing to trade speed for an inexact solution, it is reasonably easy to overcome the computational cost of bootstrapping after one estimates the penalty parameter.  For instance, we could estimate a trend filtering fit using the majorization-minimization techniques of \citet{Hunter:2005}.  An approximation to objective function~(\ref{eq:tf}), details of which are provided in Appendix~\ref{app:mm}, quickly allows us to refit trend filtering given a new set of observations $y^*$ and a pre-calculated penalty parameter, say $\hat{\lambda}_{CV(10)}$, estimated from $10$-fold cross validation.  Though this strategy is itself limited, it is quite fast.  For a fixed value of $\lambda$, it took a MacBook Pro $3.1$ GHz Intel Core i$7$ just about half the time to calculate $500$ bootstrapped approximations of $f$ in Figure~\ref{fig:boot}, with $n=50$, as it did to fit the exact solution of trend filtering and estimate $\hat{\lambda}_{CV(10)}$.  In general, we found this approximation strategy to produce estimates relatively close to the exact fit.  It was common to see mean absolute differences on the order of $10^{-3}$.  However, cross validation for trend filtering frequently produces estimates of the true function that are too ``wiggly.''  Bootstrap methods that then rely on $\hat{\lambda}_{CV}$ will also produce poor results.  The simulations and especially the real data sets presented in Section~\ref{sec:sims} highlight this point exactly.  When the estimates of $f$ under $\hat{\lambda}_{CV}$ are too wiggly, bootstrapped confidence intervals fail to contain the true function across the sampled domain, thus lowering nominal coverage probability.  

The problem with choosing $\lambda$, in our experience, largely disappears under the Bayesian approach.  Bayesian trend filtering estimates $\lambda$ by incorporating it into the hierarchical model~(\ref{eq:hier}) so that the estimates of $f$ are marginalized over all values of $\lambda$.  This encourages a more robust, stable estimate of $f$.  The results in Section~\ref{sec:sims} justify our conclusions about both trend filtering and Bayesian trend filtering.  

\section{Bayesian Trend Filtering}
\label{sec:bayes}
\subsection{Formulation}

Similar to the work of \citet{Park:2008, Kyung:2010, Griffin:2011}, and \citet{Armagan:2013}, a fully Bayesian hierarchical model for Bayesian trend filtering can be expressed as a scale mixture of normals, a mixture that was first discovered by \citet{Andrews:1974}.  The resulting hierarchical model puts a double exponential (\texttt{dexp})
 \[ 
 [f|\sigma] \propto \exp\left( -\frac{\lambda}{\sigma} ||\Dk f||_1 \right) 
 \] 
\noindent conditional prior on the penalty term $||\Dk f||_1$.  \citet{Park:2008} show that the double exponential prior conditional on $\sigma$ ensures that the joint posterior distribution of $[f,\sigma^2]$ is unimodal.  The fully Bayesian hierarchical model to fit Bayesian trend filtering is 
 
\begin{equation} 
  \label{eq:hier}
  \begin{aligned} 
    y | f, \sigma^2 & \sim \mathcal{N}_n(f, \sigma^2 I_n) \\
    f | \sigma^2, \omega_1, \ldots, \omega_{n-k-1} & \sim \mathcal{N}_n(0, \sigma^2 \Sigma_{f}), \\ 
    \Sigma_{f}^{-1} = \Sigma_{f}^{-1}(\omega_1^{-1}, \ldots, \omega_{n-k-1}^{-1}) & = (\Dk)^t \text{diag}(\omega_1^{-1}, \ldots, \omega_{n-k-1}^{-1}) \Dk \\
    \omega_1, \ldots, \omega_{n-k-1} | \lambda & \sim \prod_{j=1}^{n-k-1} \frac{\lambda^2}{2}\exp(-\lambda^2\omega_j/2)d\omega_j, \quad \omega_j > 0, \forall j \\    
    \lambda | \alpha, \rho & \sim \psi(\lambda|\alpha, \rho)d\lambda, \quad \lambda > 0 \\
    \sigma^2 & \sim \pi(\sigma^2)d\sigma^2, \quad \sigma^2 > 0. \\
  \end{aligned}
\end{equation}

\noindent The $\omega_1, \ldots, \omega_{n-k-1}$ are mutually independent, and throughout we let $\pi(\sigma^2)=\sigma^{-2}$, which is the limiting improper prior from an inverted gamma distribtion.  The distribution on $\lambda$, denoted by $\psi$, highlights the fact that a slight change of the prior will produce two different conditional prior distributions.  The more common, double exponential conditional prior is found by letting $\psi$ be a gamma distribution $\Gamma(\alpha,\rho)$ on $\lambda^2$ \citep{Park:2008,Kyung:2010}.  Additionally, we consider a variation on the prior developed by \citet{Lee:2010, Lee:2012}, and \citet{Armagan:2013} which uses a $\Gamma(\alpha,\rho)$ prior on $\lambda$ instead of $\lambda^2$.  The generalized double Pareto (\texttt{gdp}) distribution takes on the following form within Bayesian trend filtering,
\[ 
[f|\sigma] =  \frac{1}{2 \sigma\rho/\alpha}\left( 1 + \frac{1}{\alpha}\frac{||\Dk f||_1}{\sigma\rho/\alpha} \right)^{-(n-k-1+\alpha)}.
\]  
\noindent The \texttt{gdp} is known to eschew much of the bias otherwise induced by the exponential tails of the double exponential distribution \citep{Lee:2010}.  We explore the choice of hyperparameters $\alpha,\rho$ in Section~\ref{sec:sims}.  

A simplie Gibbs sampler provides samples from the posterior distributions of interest for Bayesian trend filtering.  The full conditionals shared by both conditional priors are

\begin{align*}
  [f| \cdot ] & \sim \mathcal{N}_n\left( (I_n + \Sigma_{f}^{-1})^{-1}y, \sigma^2(I_n + \Sigma_{f}^{-1})^{-1}\right), \\
  [1/\omega_j | \cdot ] & \sim \text{inverse-Gaussian} \left( \sqrt{\frac{\lambda^2 \sigma^2}{|(\Dk f)_j|^2}}, \lambda^2  \right), \quad \forall j, \\
  [\sigma^2 | \cdot ] & \sim \Gamma^{-1} \left( n, \frac{1}{2}(y - I_nf)^t(y - I_nf) + \frac{1}{2} f^t\Sigma_{f}^{-1}f  \right). \\
\end{align*}
\noindent The full conditional of $\lambda^2$ for \texttt{dexp} is $\Gamma(n-k-1+\alpha, \sum_{j=1}^{n-k-1}\omega_j/2 + \rho)$, and relative to \texttt{gdP}, $[\lambda|\cdot] \sim \Gamma(n-k-1+\alpha, ||\Dk f||_1 / \sigma + \rho)$.

Model (\ref{eq:hier}) can be viewed as an extension of the work in \citet{Kyung:2010}.  Therefore, we appeal to their Propositions $4.1$ and $4.2$ which show that the underlying Gibbs sampler is geometrically ergodic.  The two Gibbs samplers, for \texttt{dexp} and \texttt{gdp}, converge both in theory and in practice very quickly. 

\begin{proposition}
  \label{theorem:ergodic}
  The Gibbs sampler for the hierarhcical model (\ref{eq:hier}) is geometrically ergodic.
\end{proposition}

\noindent We refer the reader to the proof by \citet{Kyung:2010}.  In Section~\ref{sec:discussion}, we show that Proposition~\ref{theorem:ergodic} can help reduce the computational cost of Bayesian trend filtering.

\begin{figure}[H]
  \centering
  \includegraphics{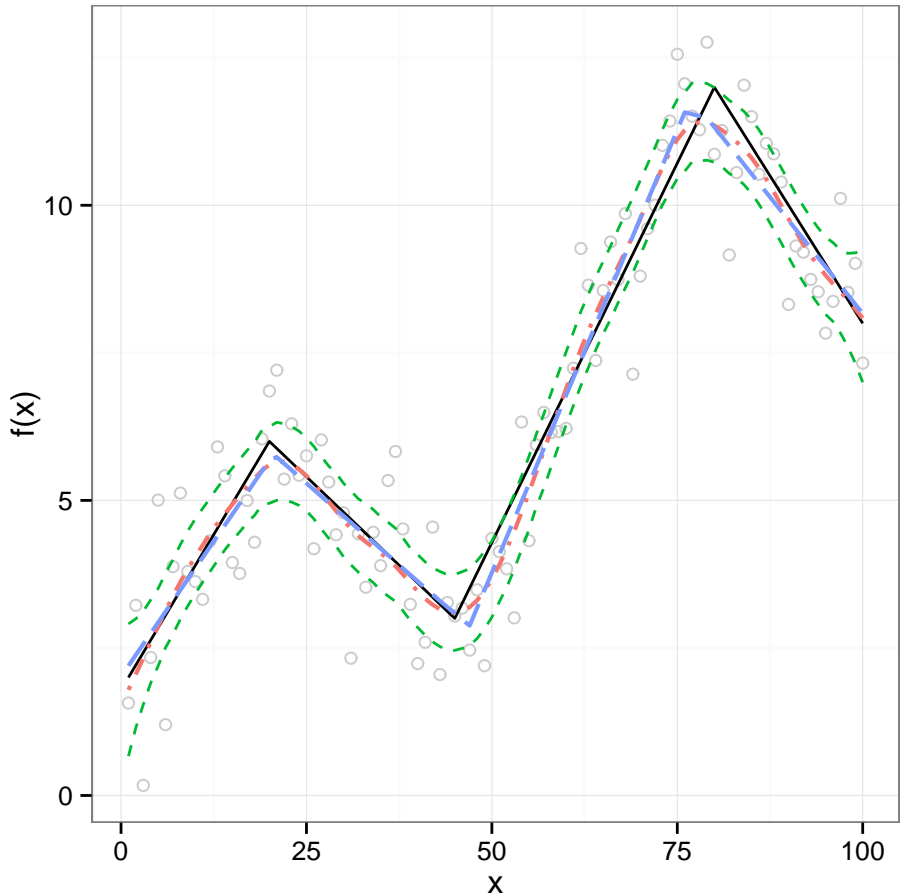}
  \caption{A piecewise linear function (solid black) with knots at $20, 45$, and $80$ fit by Bayesian trend filtering (red dot-dash) and the original trend filtering (purple dash).  $95$\% credible intervals for Bayesian trend filtering (green dash) cover the true function and the frequentist trend filtering fit across the entire domain.}
  \label{fig:linear}
\end{figure}

A few points contrasting trend filtering with Bayesian trend filtering should be noted.  Trend filtering, by restricting the parameter space of the objective function (\ref{eq:tf}), automatically sets some terms in the penalty to exactly zero.  Such a data dependent selection of important predictors is philosophically appealing.  Within trend filtering, this data dependent selection corresponds to setting esimates of the terms in the total variation of $f^{(k)}$ to zero.  Bayesian trend filtering, however, never sets any terms identically to zero.  Figure~\ref{fig:linear} compares Bayesian trend filtering to the original trend filtering formulation.  Bayesian trend filtering doesn't quite predict a piecewise linear function when in fact the true function is a piecewise linear function, with three knots at $x$ equal to $20, 45$, and $80$.  What Bayesian trend filtering sacrifices in knot detection, it makes up for when fitting smooth curves.  The information gained by incorporating $\lambda$ into the Gibbs sampler, and the propagation of that information back to the estimates of $f$ provides stable estimation, as is shown in Section~\ref{sec:ex}.  Thus, the primary advantage of Bayesian trend filtering is as a smoother and not as a knot-detection method.

\subsection{Computational Considerations}
\label{sec:computation}

The very idea of the (generalized) lasso, to shrink some of the elements of the penalty term towards zero possibly setting some to exactly zero, can cause computational issues in the Bayesian setting.  Consider the full conditional $[1/\omega_j|\cdot]$, for either conditional prior \texttt{dexp} or \texttt{gdp}.  The mean in the inverse-Gaussian distribution inverts exactly that which we are shrinking towards zero.  Any sample from the posterior distribution such that an element of the penalty term is very close to zero, threatens numerical stability when drawing samples from the already rather numerically sensitive inverse-Gaussian distribution; see \citep{Wheeler:2013} and the references there within.  With Bayesian trend filtering, when an element of $|\Dk f|$ is too small, simulating a draw from $[1/\omega_j|\cdot]$ can return a value less than or equal to zero -- obviously a problem for a distribution with non-negative support.  To ameliorate such issues, we propose the admitedly inelegant solution of resampling the entire vector $f$ if any element of $|\Dk f|$ is less than $10^{-10}$.  We find that resampling happens less than five percent of time.  Further, when resampling does occur, it is extremely rare that more than one resample is ever needed.  Despite such numerical issues, restricting the support of specific posterior distributions does not appear to hinder Bayesian trend filtering's performance; see Section~\ref{sec:sims}.

\section{Numerical Results}
\label{sec:ex}
\subsection{Simulations}
\label{sec:sims}

We compare Bayesian trend filtering (\texttt{BTF}), with both conditional priors \texttt{dexp} and \texttt{gdp}, against three different methods: trend filtering (\texttt{TF}) from \citet{Tibshirani:2014}, Bayesian additive regression trees (\texttt{BTree}) from \citet{Chipman:2010, Kapelner:2013}, and cubic smoothing splines \texttt{CSM} from \citet{Wood:2006}.  \texttt{BTree} is included as it is becoming an increasingly popular Bayesian regression method.  We include smoothing splines as they are arguably the most used method of smoothing, and also to highlight a different point than was made of the same comparison by \citet{Tibshirani:2014}.  There, a strong argument was made for the efficiency of \texttt{TF}, implicitly defined to be mean squared error (mse) per degree of freedom.  In that world, \texttt{TF} clearly stands above as its asymptotic results are shown to benefit finite sample sizes.  Here, a more applied world is hypothesized, where estimation of $\lambda$ further dictates a methods performance.  

A functions used in the simulations are from various \texttt{R} \citep{R:2014} packages.  \texttt{BTree} was fit with \texttt{bartMachine::bartMachine} using the default values of $50$ trees and $9000$ (after burn-in) iterations \citep{Kapelner:2013}.  \texttt{CSM} was fit using the cubic smoothing spline function \texttt{mgcv::gam}, with all inputs used as knots (to make more fair the comparison between \texttt{CSM} and the trend filtering methods).  \texttt{TF} was fit with \texttt{genlasso::trendfilter} using a cubic piecewise polynomial, with both $5$- and $10$-fold cross validation \citep{Tibshirani:2014}.  \texttt{BTF}, also using a cubic piecewise polynomial, was fit using \texttt{btf:btf} with $9000$ (after burn-in) posterior samples \citep{Roualdes:2014}.  Since the choice of hyperparameters $\alpha$ and $\rho$ effects the overall fit of both \texttt{BTF-gdp} and \texttt{BTF-dexp}, we explore Bayesian trend filtering's responsiveness to changes in these hyperparameters.  We fit \texttt{BTF-dexp} with $\alpha  \in \{0.1, 0.5, 1, 1.5, 2\}$ and $\rho \in \{10^{-4}, 10^{-3}, 10^{-2}, 0.1, 1\}$, and \texttt{BTF-gdp} with each of $\alpha \in \{0.5, 0.1, 1, 1.5, 2\}$ and $\rho \in \{10^{-3}, 10^{-2}, 0.1, 1\}$.  We chose a number of ``small'' values of $\rho$ to ensure a true thresholding rule, and thus encourage shrinkage \citep{Armagan:2013}.  

For simulated data, we consider two univariate functions $f: [0,1] \mapsto \mathbb{R}$ with $100$ regularly spaced inputs.  The first, a piecewise cubic function is borrowed from \citet{Tibshirani:2014}.  The second is a difficult, spatially inhomogeneous function, colloquially known as dampened harmonic motion (dhm).  $1000$ replications of these two functions with three different levels of normal noise are evaluated upon three criteria.  For each replication we calculate the mean and standard deviation of the mean squared errors (mse) and $95\%$ confidence intervals for both the underlying function evaluated at all inputs and the variance.  $1000$ bootstrap samples were used to estimate confidence intervals of function estimates for \texttt{TF}, while \texttt{CSM} used the method \texttt{predict.gam} to obtain standard errors.  All frequentist methods relied on the bootstrap to estimate $\sigma^2$.  We used posterior samples to create credible intervals for all the Bayesian methods.  Thus, the $1000$ replications are used to estimate the mean and standard deviation of the mses, and overall mean and standard errors of the two different coverage probabilities.  

The real world data consists of two data sets common to the smoothing literature.  The first is a dataset of global mean surface temperature deviations for the years $1881$ to $2005$ from \citet{Hodges:2013}.  The second is the SILSO dataset, monthly sunspot counts for the years $[1980, 2014]$.

\subsubsection{Piecewise Cubic}

\begin{figure}[H]
  \centering
  \begin{subfigure}[b]{0.48\textwidth}
    \includegraphics[width=\textwidth]{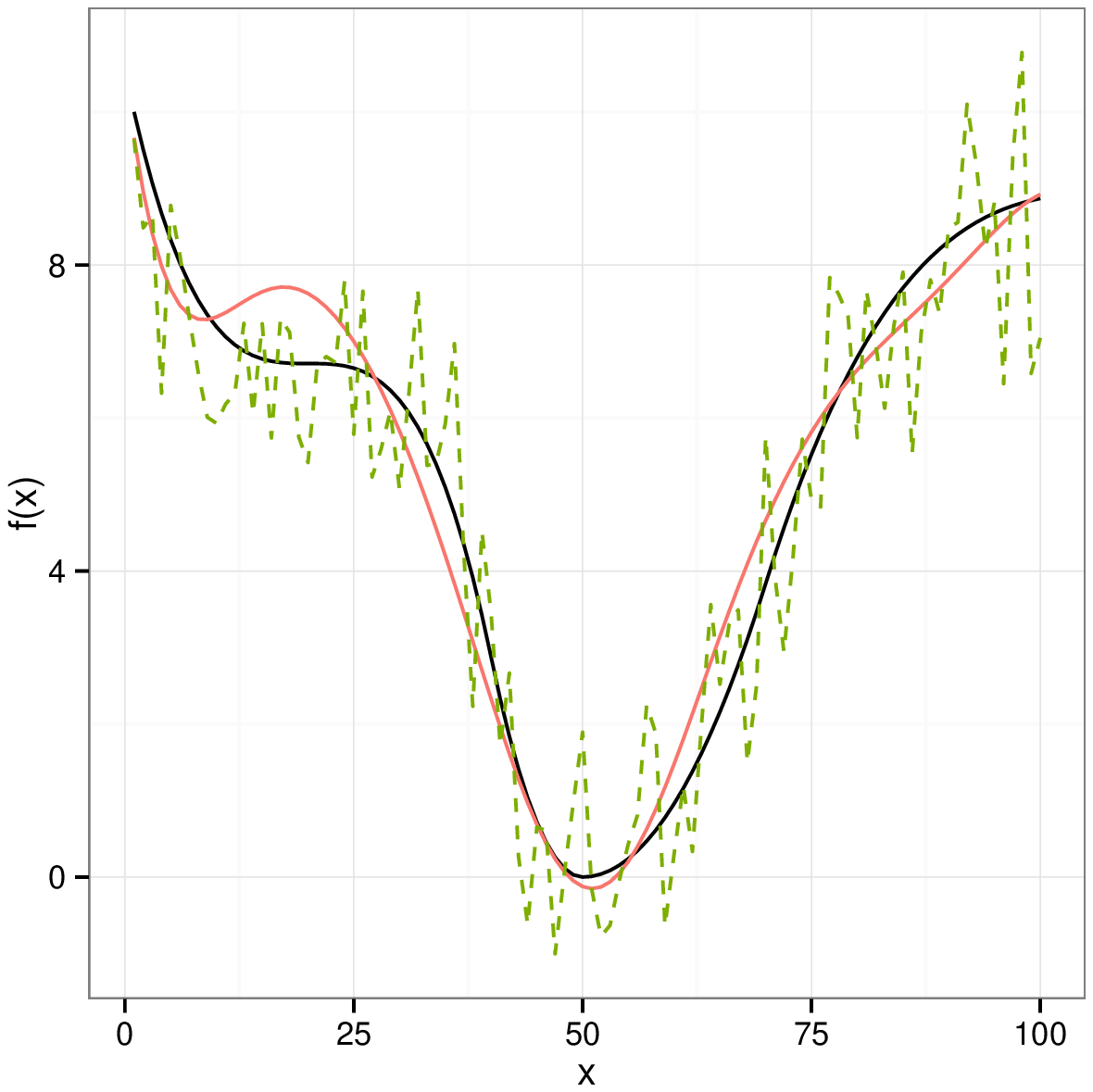}
  \end{subfigure}
  \begin{subfigure}[b]{0.48\textwidth}
    \includegraphics[width=\textwidth]{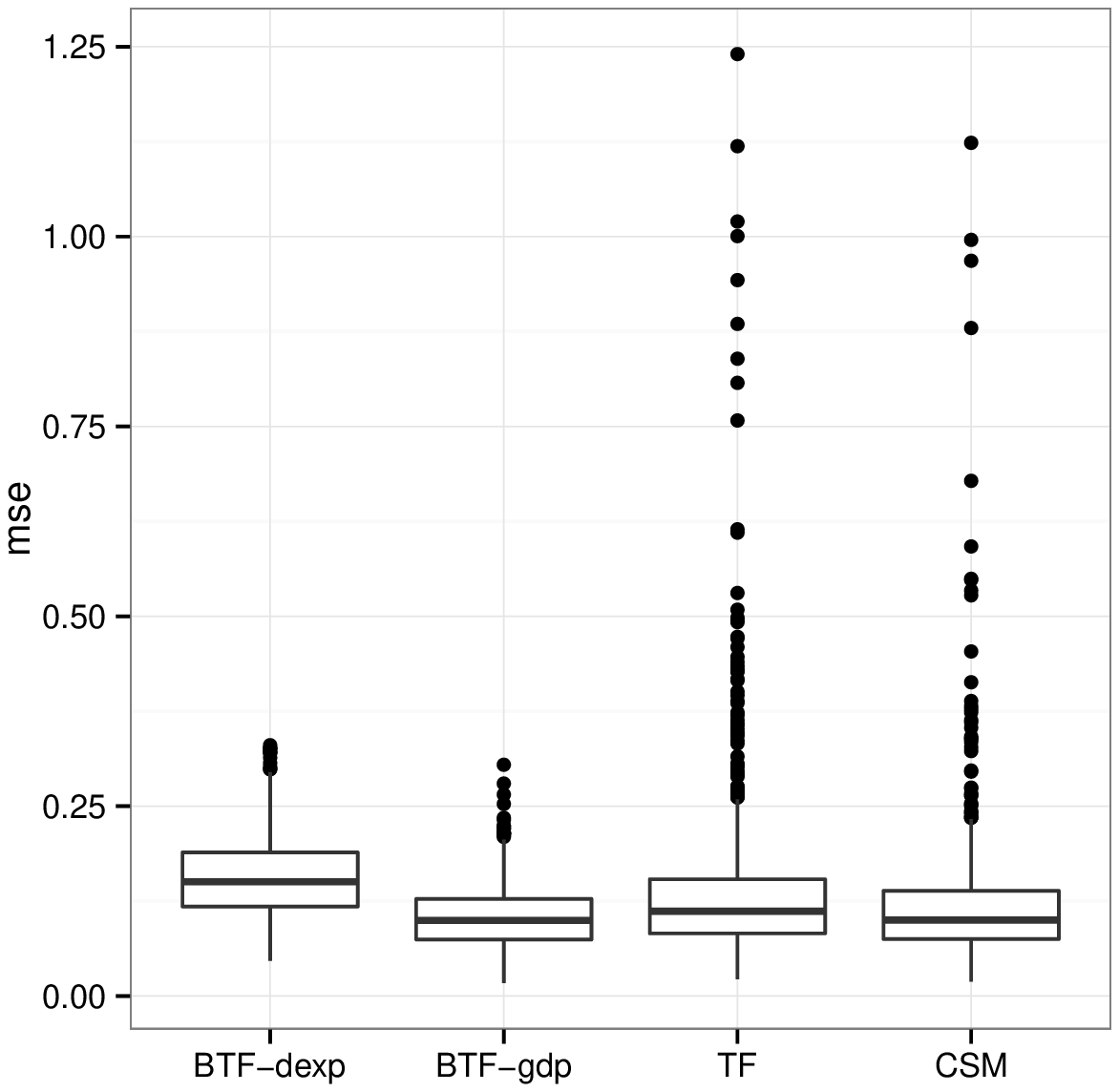}
  \end{subfigure}
  \caption{Both plots use noise level $\sigma:=1$ and the \texttt{BTF} methods use $\alpha:=1$ and $\rho:=10^{-2}$.  Left: The true piecewise cubic function (solid black) and the worst fits for the methods \texttt{BTF-gdp} (solid red) and \texttt{CSM} (dash green).  Right: All $1000$ replications of mses computed from fits of the labeled methods to the piecewise cubic function.}
  \label{fig:pwc}
\end{figure}

Of the three levels of $\mathcal{N}(0, \sigma^2)$ noise, $\sigma \in \{0.75, 1.0, 1.25\}$, for the piecewise cubic function with $\sigma := 1.0$, the right plot of Figure~\ref{fig:pwc} shows the mses of $1000$ replications of $100$ observations from the true function (solid black) displayed in the left plot.  The \texttt{BTF} methods displayed in this plot used the hyperparameter values of $\alpha:=1$ and $\rho:=10^{-2}$.  These plots are representative of all cases of the \texttt{BTF} methods for which $\rho<1$, inclusive of all the tested values of $\alpha$.  \texttt{BTF} clearly produces mses with lower variance, although most methods save \texttt{BTree} (not shown) provide practically the same median mse.  Still, \texttt{BTF-gdp} in Figure~\ref{fig:pwc} provided nearly the smallest set of mses: there are $45$ and $25$ fits from \texttt{TF} and \texttt{CSM}, respectively, where the mse is greater than the largest mse for \texttt{BTF-gdp}.  And conversely, there are no fits that provide an mse smaller than the smallest mse from \texttt{BTF-gdp}.  Table~\ref{tab:pwc_mse} provides mean and standard deviations of the mses for the different methods across all levels of noise chosen for the piecewise cubic function.

The left plot of Figure~\ref{fig:pwc} displays the fits with the greatest mse of \texttt{BTF-gdp} and \texttt{CSM}.  The \texttt{BTF} methods appear to minimize the worst of the fits to the piecewise cubic function.   We highlight \texttt{BTF-gdp} because it almost unanimously outperforms \texttt{BTF-dexp} in all of our examples and we highlight \texttt{CSM} because it is the next best method, outside of the \texttt{BTF} methods, in terms of methods with extreme mse values.

\begin{table}[ht]
\centering
\begin{tabular}{lcccccc}
  & \multicolumn{2}{c}{$\sigma=0.75$} & \multicolumn{2}{c}{$\sigma=1$} & \multicolumn{2}{c}{$\sigma=1.25$} \\
\hline
method & mean & sd & mean & sd & mean & sd \\ 
  \hline
  BTF-dexp & $8.9$  & $3.0$ & $16$ & $5.2$  & $25$ & $7.9$ \\
  BTF-gdp  & $\mathbf{6.6}$  & $\mathbf{2.4}$ & $\mathbf{11}$ & $\mathbf{4.2}$  & $\mathbf{15}$ & $\mathbf{6.4}$ \\
  TF       & $7.6$  & $5.9$ & $14$ & $11$  & $20$ & $17$ \\ 
  CSM      & $7.2$  & $5.3$ & $12$ & $8.6$  & $18$ & $14$ \\
  \hline
\end{tabular}
\caption{Mean and standard deviations of the $1000$ mean squared errors for each of the three noise levels, rounded and multiplied by $100$ for ease of comparison.  The smallest value(s) within each column is(are) bold.  The \texttt{BTF} methods used the hyperparameters $\alpha:=1$ and $\rho:=10^{-2}$.}
\label{tab:pwc_mse}
\end{table}

\begin{figure}[H]
  \centering
  \begin{subfigure}[b]{0.48\textwidth}
    \includegraphics[width=\textwidth]{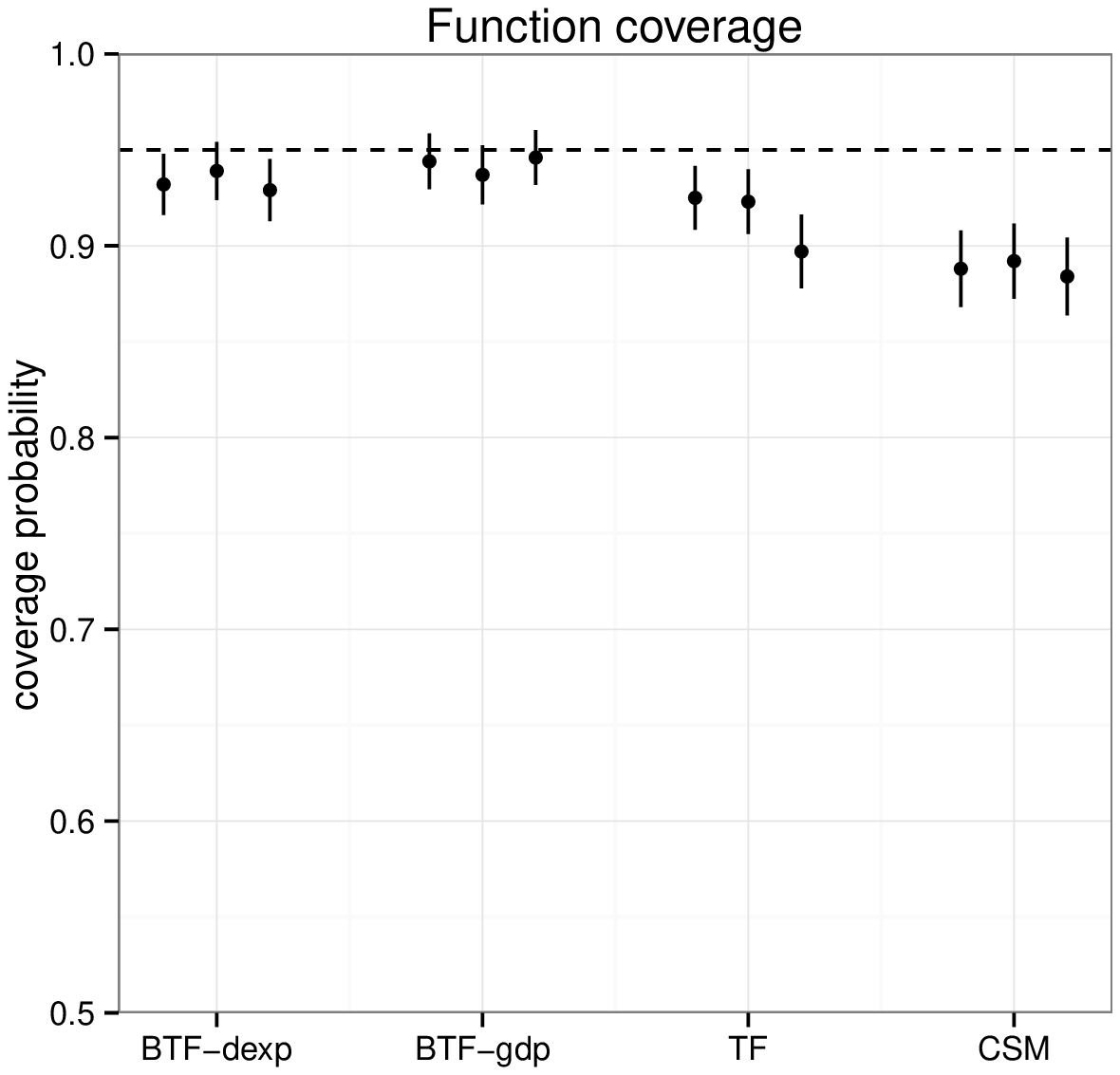}
  \end{subfigure}
  \begin{subfigure}[b]{0.48\textwidth}
    \includegraphics[width=\textwidth]{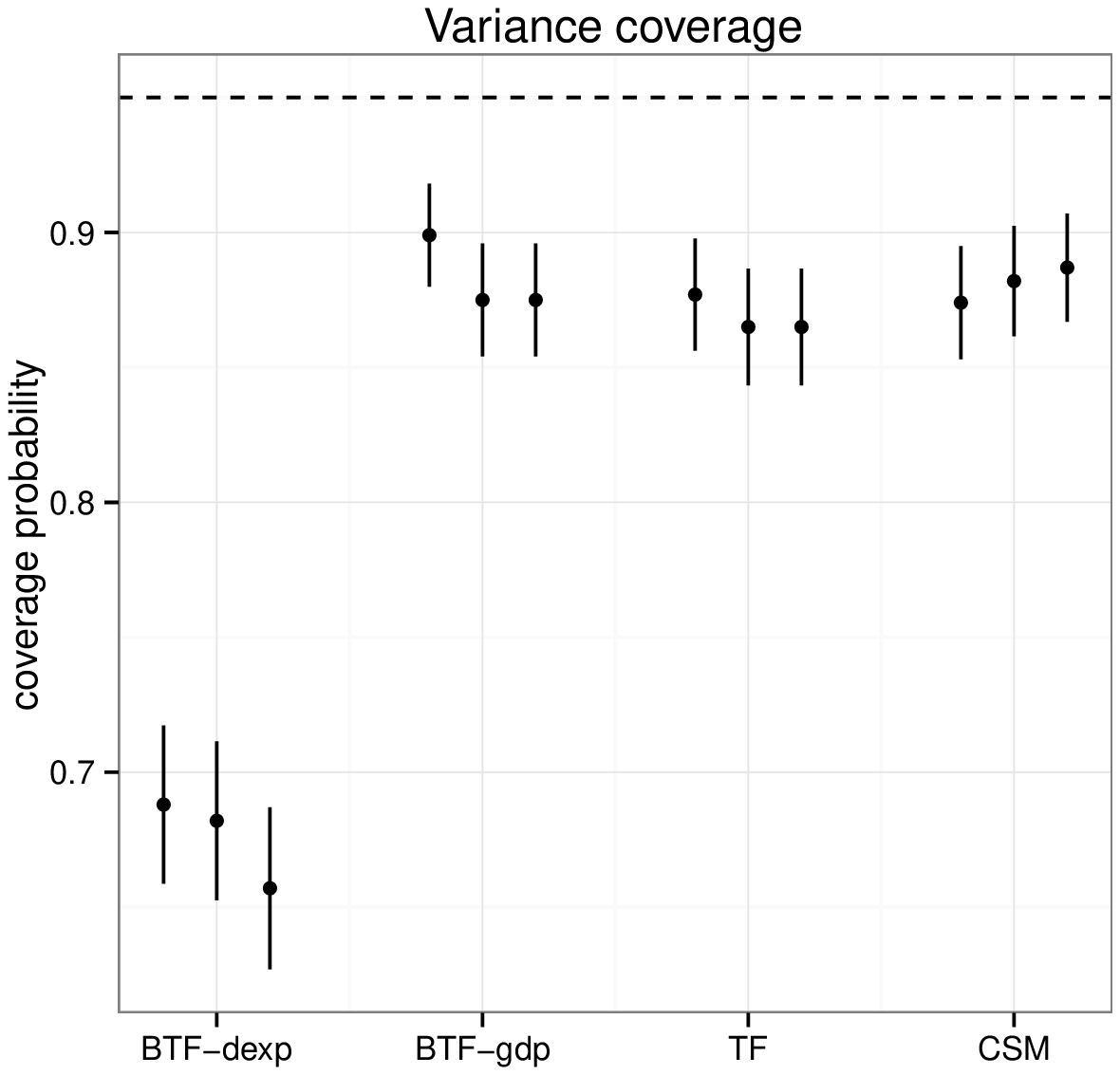}
  \end{subfigure}
  \caption{Each method's function and variance coverage probablities are displayed for every level of noice considered for the piecewise cubic function.  The \texttt{BTF} methods used the hyperparameter values $\alpha:=1$ and $\rho:=10^{-2}$.}
  \label{fig:pwc_cov}
\end{figure}

Consistent function estimation by \texttt{BTF} is seen when we compare coverage probabilities.  Figure~\ref{fig:pwc_cov} plots the mean plus/minus two standard errors, at every level of noise, of function and variance coverage probabilities.  Since we measured function coverage across the entire domain, the over-fitting of \texttt{CSM} and \texttt{TF} reduced their respective function coverage probabilities.  Moreover, no method appropriately covered the variance at the nominal value.  We notice, though do not provide plots of such here, that both \texttt{BTF} methods' variance coverage declines as the hyperparameter $\rho$ increases.

While \texttt{BTree} (not shown) provides reasonably small mses for the piecewise cubic function, it had the highest median mse.  This is largely due to the fact that it is not a smoothing technique.  \texttt{BTree} was removed from the plots throughout as it distracted from the main comparisons of interest.  Still, \texttt{BTree} did quite well estimating the variance.  In other contexts, where a smooth fit is not necessary, \texttt{BTree} has much to offer beyond what most of these smoothing techniques are able to handle.  

\subsubsection{Spatially Inhomogeneous}

\begin{figure}[H]
  \centering
  \begin{subfigure}[b]{0.48\textwidth}
    \includegraphics[width=\textwidth]{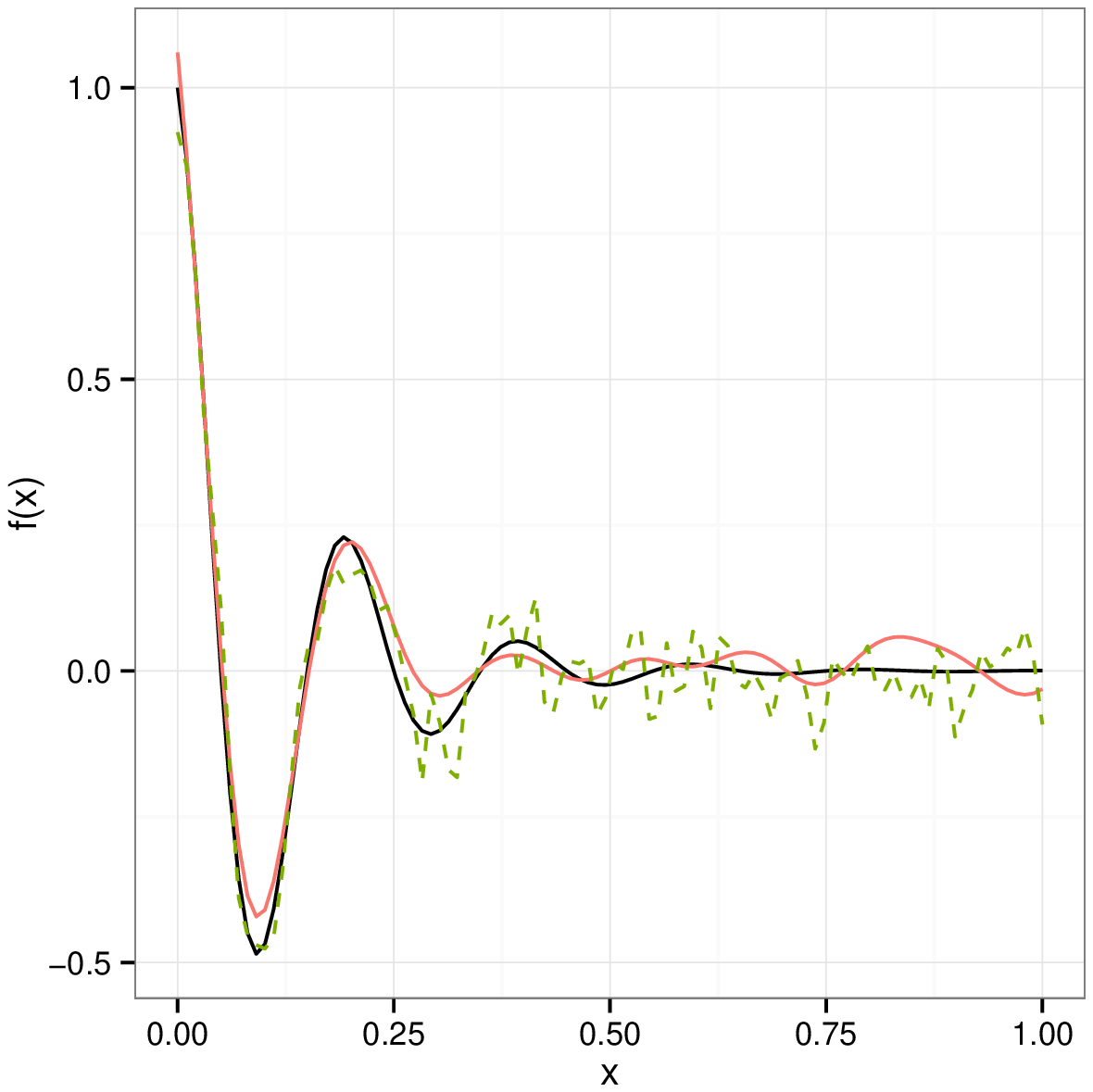}
  \end{subfigure}
  \begin{subfigure}[b]{0.48\textwidth}
    \includegraphics[width=\textwidth]{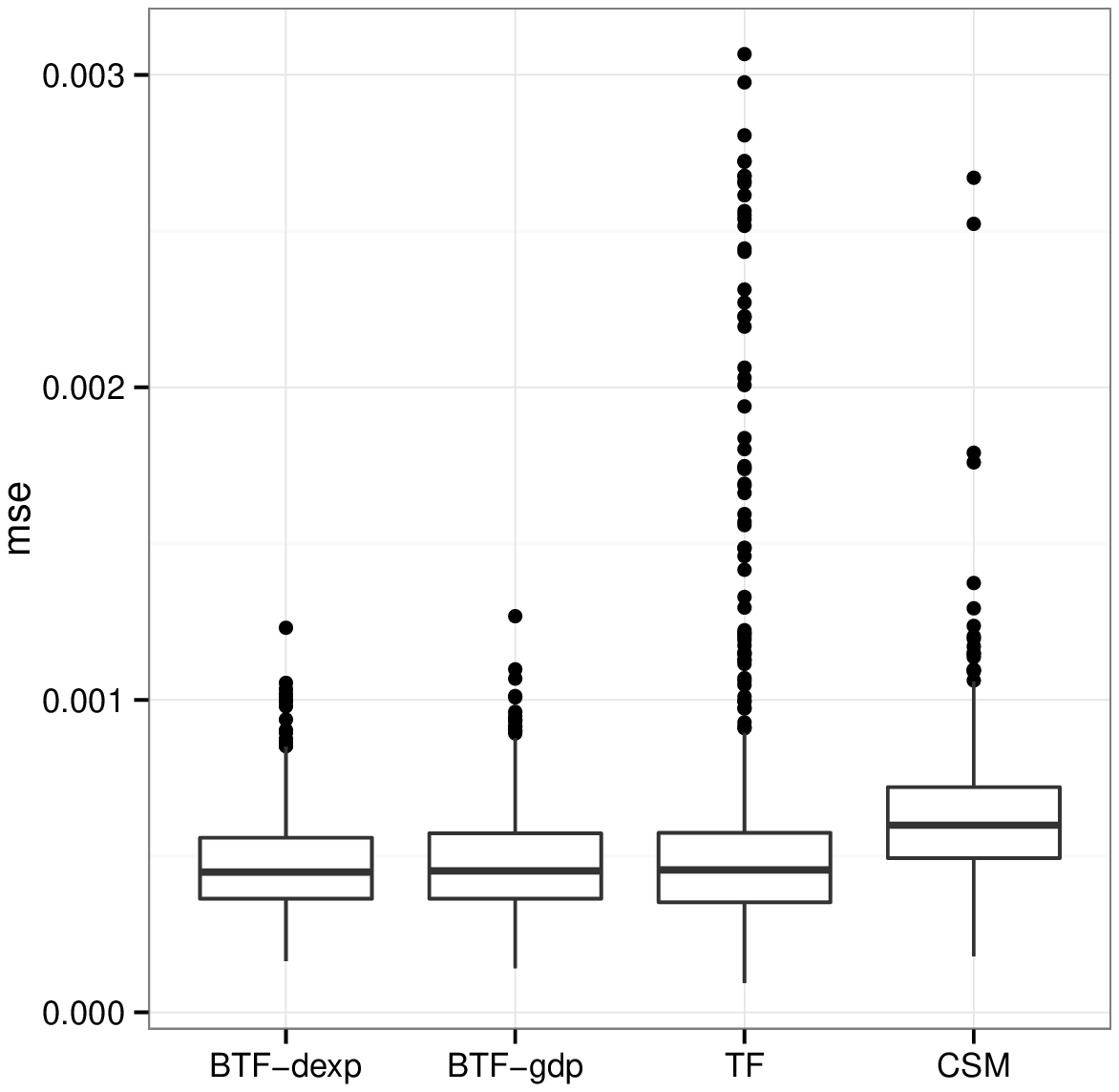}
  \end{subfigure}
  \caption{Both plots use noise level $\sigma:=0.05$ and the \texttt{BTF} methods use $\alpha:=1$ and $\rho:=10^{-2}$.  Left: The true dampened harmonic motion function (solid black) and the worst fits for the methods \texttt{BTF-gdp} (solid red) and \texttt{CSM} (dash green).  Right: All $1000$ replications of mses computed from fits of the labeled methods to the dampened harmonic motion function.}
  \label{fig:dhm}
\end{figure}

The second example uses the spatially inhomogeneous function \[  f(x) = \exp\{-7.5x\}\cos(10\pi x),\] with $\mathcal{N}(0, \sigma^2)$ noise and $\sigma \in \{0.025, 0.05, 0.075\}$.  Figure~\ref{fig:dhm} again plots the true function (solid black) on the left and all $1000$ mses on the right, for the noise level $\sigma:=0.05$.  The \texttt{BTF} methods presented for dampened harmonic motion use the same hyperparameter values as for the piecewise cubic function, $\alpha:=1$ and $\rho:=10^{-2}$.  Like with the piecewise cubic function, these hyperparameter values generally represent the all hyperparameter choices where $\rho<1$.   In Figure~\ref{fig:dhm} we see that all trend filtering methods produce nearly identical median mses.  The over-fitting is still a problem for \texttt{CSM} and \texttt{TF}.  Also, notice that the median mse for the cubic smoothing spline method \texttt{CSM} is a bit larger than the median mses for the trend filtering methods.  Table~\ref{tab:dhm_mse} summarizes the mses for all levels of noise for the dampened harmonic motion function.  The \texttt{BTF} methods provide the smallest mean mses, with the smallest standard deviation, for all levels of noise considered.

Generally with the dampened harmonic motion, as judged by mses, the \texttt{BTF} methods again perform well.  In this case, \texttt{BTF-dexp} performs insignificantly better than \texttt{BTF-gdp}.  This could be due to the heavy smoothing, required for dapmened harmonic motion, that the exponential tails of \texttt{dexp} encourage.  Still, \texttt{BTF-gdp} gives smaller mean mses with smaller variation than the other methods.  Greater than the largest mse of \texttt{BTF-gdp} there exists $42$ and $6$ mses as produced by \texttt{TF} and \texttt{CSM}, respectively.  Only \texttt{TF} provides any, specifically two, simulations with a smaller mse than that of \texttt{BTF-gdp}.

\begin{table}[ht]
\centering
\begin{tabular}{lcccccc}
  & \multicolumn{2}{c}{$\sigma=0.025$} & \multicolumn{2}{c}{$\sigma=0.05$} & \multicolumn{2}{c}{$\sigma=0.075$} \\
\hline
method & mean & sd & mean & sd & mean & sd \\ 
  \hline
  BTF-dexp & $\mathbf{1.3}$ & $\mathbf{0.39}$ & $\mathbf{4.7}$ & $\mathbf{1.5}$ & $\mathbf{9.9}$ & $\mathbf{3.1}$ \\
  BTF-gdp  & $\mathbf{1.3}$ & $\mathbf{0.39}$ & $4.8$ & $1.6$ & $11$ & $3.9$ \\
  TF       & $1.5$ & $0.92$ & $5.4$ & $3.9$ & $11$ & $7.2$ \\ 
  CSM      & $1.9$ & $0.53$ & $6.2$ & $2.0$ & $12$ & $4.2$ \\
  \hline
\end{tabular}
\caption{Mean and standard deviations of the $1000$ mean squared errors for each of the three noise levels of the dampened harmonic motion function, rounded and multiplied by $10^3$ for ease of comparison.  The smallest value(s) within each column is(are) bold.  The \texttt{BTF} methods used the hyperparameter values $\alpha:=1$ and $\rho:=10^{-2}$.}
\label{tab:dhm_mse}
\end{table}

\begin{figure}[H]
  \centering
  \begin{subfigure}[b]{0.48\textwidth}
    \includegraphics[width=\textwidth]{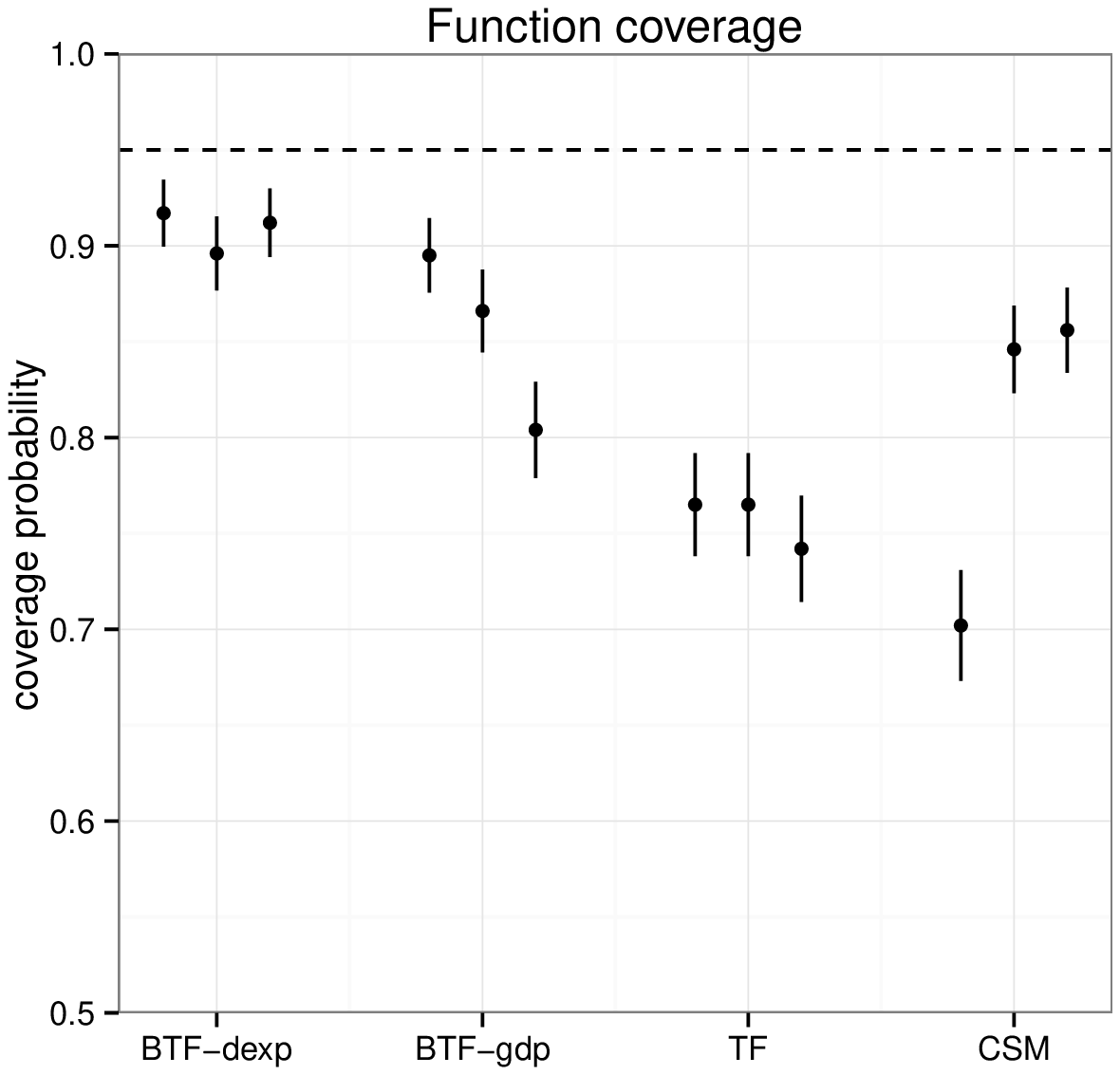}
  \end{subfigure}
  \begin{subfigure}[b]{0.48\textwidth}
    \includegraphics[width=\textwidth]{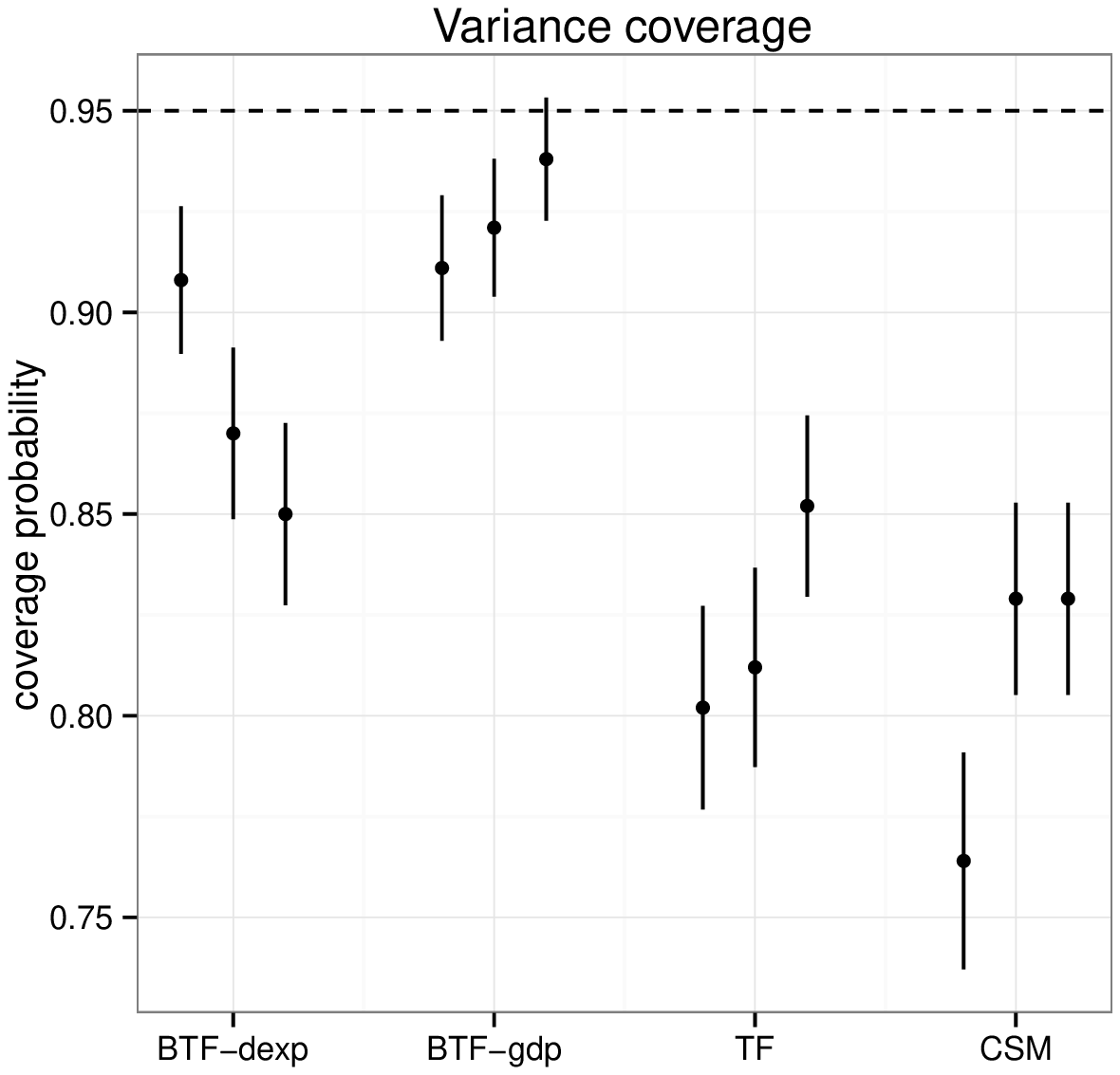}
  \end{subfigure}
  \caption{Each method's function and variance coverage probablities are displayed for every level of noise considered for the dampened harmonic motion function.  The \texttt{BTF} methods used the hyperparameter values $\alpha:=1$ and $\rho:=10^{-2}$.}
  \label{fig:dhm_cov}
\end{figure}

Figure~\ref{fig:dhm_cov} shows function and variance coverage probability means plus/minus two standard errors for the all levels of noise of the dampened harmonic motion function.  Function estimation proved difficult for all the methods.  This is likely do to the fact that we are measuring overall function coverage on a quite spatially inhomogeneous function.  Still for many values of the hyperparameters, the \texttt{BTF} methods perform well in terms of function coverage probability, as is seen in the left plot of Figure~\ref{fig:dhm_cov}.  As for variance coverage, \texttt{BTF-dexp} does noticeably worse than \texttt{BTF-gdp}, which itself outperforms the other methods.

\subsection{Real Data Analysis}
\label{sec:real}
\subsubsection{Surface Temperatures}

\begin{figure}[H]
  \centering
  \begin{subfigure}[b]{0.48\textwidth}
    \includegraphics[width=\textwidth]{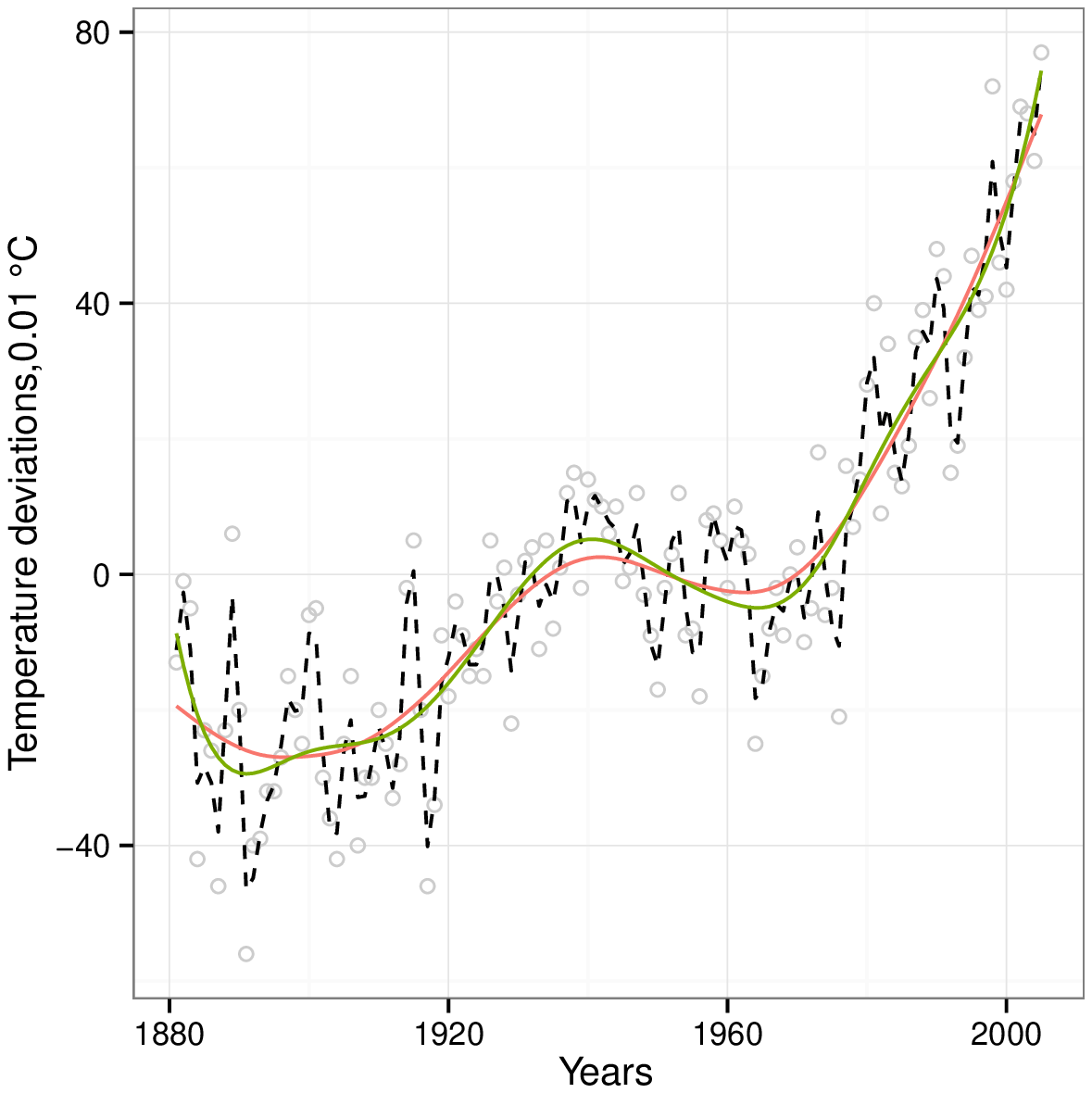}
  \end{subfigure}
  \begin{subfigure}[b]{0.48\textwidth}
    \includegraphics[width=\textwidth]{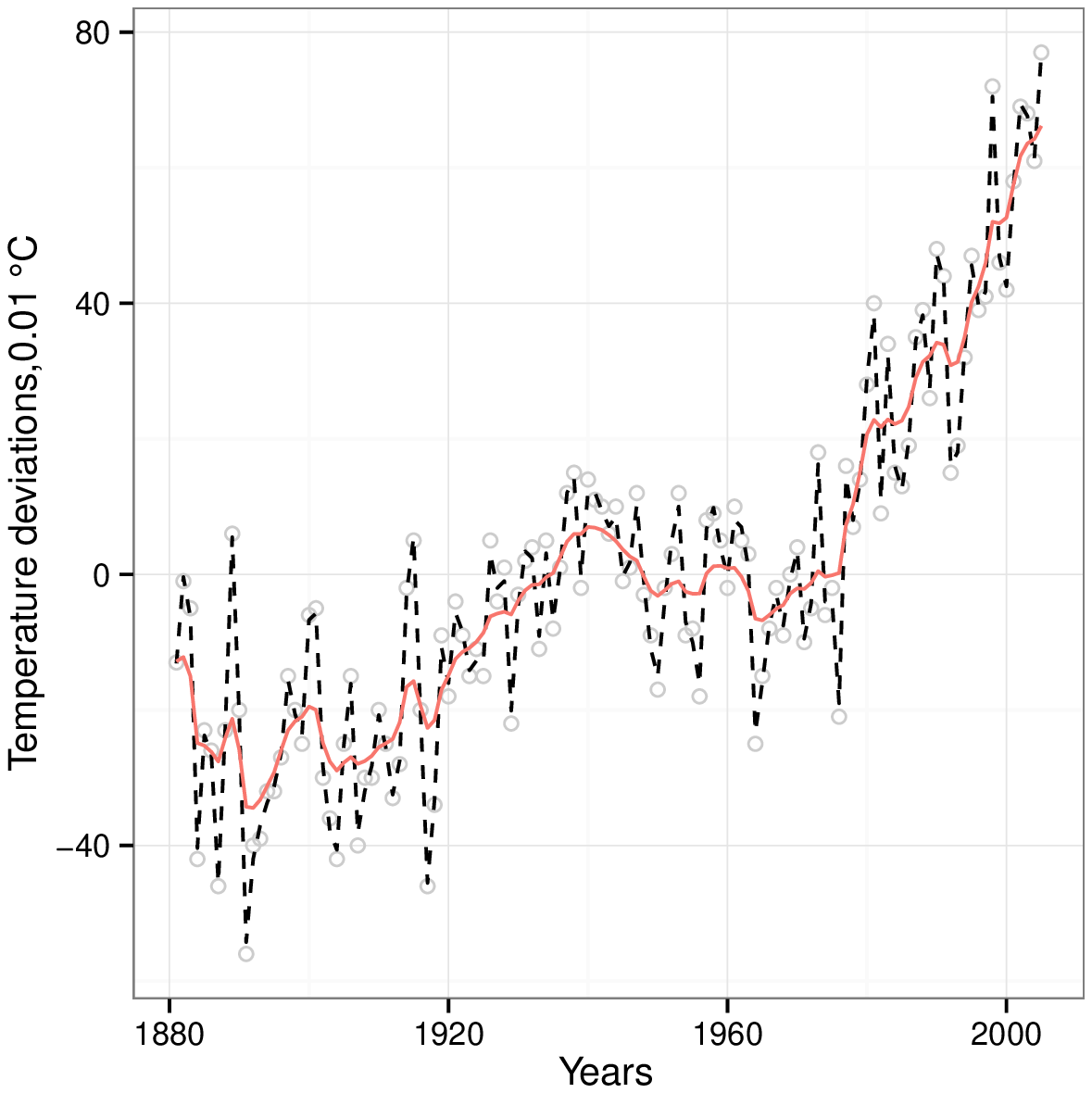}
  \end{subfigure}
  \caption{Plotted are the $125$ annual, global mean surface temperature deviations from years $1881$ until $2005$, inclusive \citep{Hodges:2013}.  Left: The fits of \texttt{CSM} (dash black), \texttt{CSM} with an AR(1) error structure (solid red), and \texttt{BTF-gdp} with $\alpha:=1$ and $\rho:=10^{-2}$ (solid green) are displayed.  Right: Two estimates are shown, \texttt{BTree} (solid red) and \texttt{TF} using $10$-fold cross validation (dash black).}
  \label{fig:global}
\end{figure}

The global surface temperature deviations data consist of yearly, $[1881, 2005]$, measurements in units of $0.01\,^{\circ}\mathrm{C}$.  Figure~\ref{fig:global} displays various fits to these data.  The global surface temperature data proves to be a problem for both $5/10$-fold cross validation used by \texttt{TF} (dash black), as is seen in the right plot of Figure~\ref{fig:global}.  Trend filtering essentially fits every data point.  \texttt{BTF-gdp} (solid green), where $\alpha:=1$ and $\rho:=10^{-2}$, smooth these data toward the hypothesized underlying function as seen in the left plot of Figure~\ref{fig:global}.  Because the trend filtering methods assume independent observations, possibly an incorrect error strucutre for these data, we also fit \texttt{CSM} with an autoregressive one, denoted AR(1), error structure (solid red).  We find that the estimates of \texttt{BTF-gdp} and \texttt{CSM} with an AR(1) error structure, displayed in the left plot of Figure~\ref{fig:global}, are quite similar.  In fact, \texttt{BTF-gdp}'s credible intervals (not shown) completely contain the \texttt{CSM} autoregressive one fit.

\subsubsection{Sunspots}

\begin{figure}[H]
  \centering
  \begin{subfigure}[b]{0.48\textwidth}
    \includegraphics[width=\textwidth]{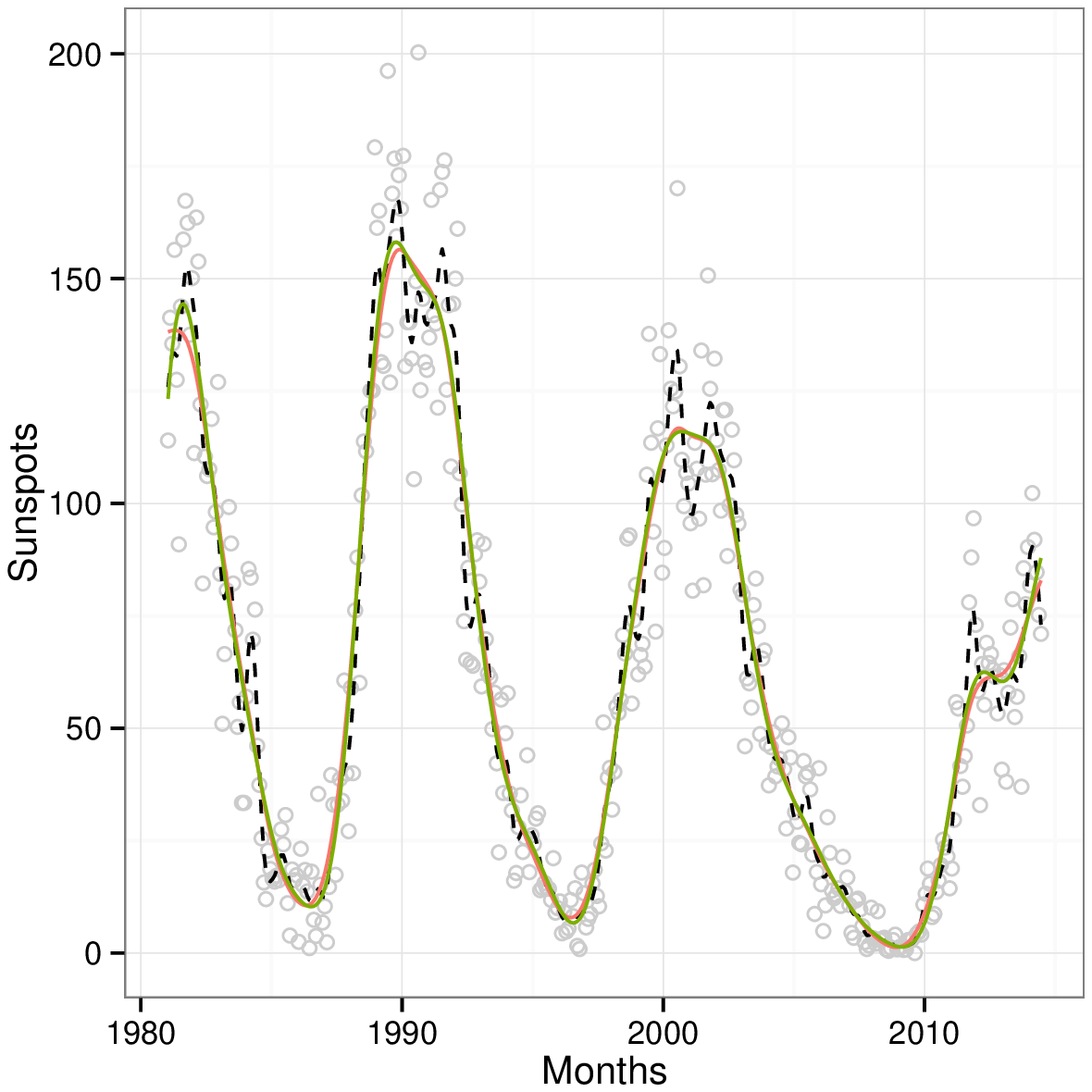}
  \end{subfigure}
  \begin{subfigure}[b]{0.48\textwidth}
    \includegraphics[width=\textwidth]{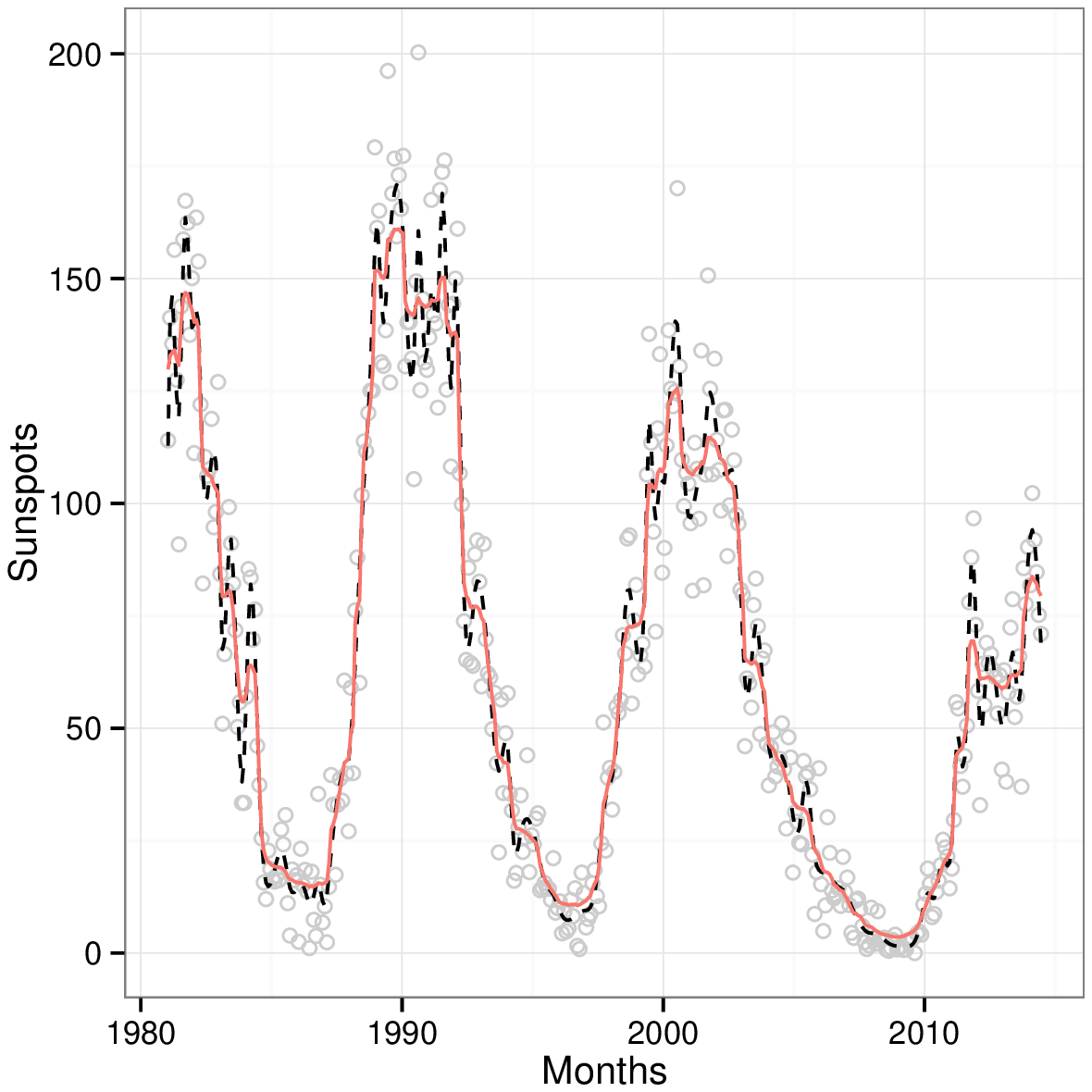}
  \end{subfigure}
  \caption{Plotted are the $402$ monthly sunspot observations from $1980$ to $2014$ \citep{SILSO:1980}.  Left: The fits of \texttt{CSM} (dash black), \texttt{CSM} with an AR(1) error structure (solid red), and \texttt{BTF-gdp} with $\alpha:=1$ and $\rho:=10^{-2}$ (solid green) are displayed.  Right: Two estimates are shown, \texttt{TF} (dash black) with $5$-fold cross validation and \texttt{BTree} (solid red).}
  \label{fig:sun}
\end{figure}

The second applied data set we investigate is monthly sunspot counts over the years $1980$ to $2014$, with $402$ observations \citep{SILSO:1980}.  In the left plot of Figure~\ref{fig:sun}, \texttt{BTF-gdp} (solid green) with $\alpha:=1$ and $\rho:=10^{-2}$ provides a nice smooth fit to these data.  In the right plot of Figure~\ref{fig:sun}, we see that \texttt{TF} (dash black) using $5$-fold cross validation provides a noisy fit to these data.  Both $5$- and $10$-fold cross validation provide nearly the same fit to these data.  \texttt{BTree} (solid red), in the left plot, provides a slightly less noisy estimate of these data than did \texttt{TF}.

Similar concerns about a possibly misspecified error structure exist for these data when applying the trend filtering methods.  We therefore fit \texttt{CSM} with an AR(1) error structure (solid red), as seen in the left plot of Figure~\ref{fig:sun}.  As with the temperature data, the \texttt{BTF-gdp} (solid green) and \texttt{CSM}-AR(1) fits are quite similar, again the estiamtes of \texttt{CSM}-AR(1) are completely contained by the credible intervals (not shown) of \texttt{BTF-gdp}.  It is unclear which assumption is more appropriate for the underlying physical process of these data: correlated errors as in \texttt{CSM} with an AR(1) error structure, or a function with evaluations correlated across time and uncorrelated errors, as Bayesian trend filtering assumes.

\section{Discussion}
\label{sec:discussion}

We developed a Bayesian, nonparametric smoother that finds its origins in the lasso literture.  Our method uses a simple scale mixture of normals to build a fully Bayesian hierarchical model.  The hierarchical model of Bayesian trend filtering closely resembles the structure of Gaussian process regression, albeit with a unique covariance function that is best understood in terms of the underlying penalty; see \citet{Rasmussen:2006} for a thorough review of Gaussian processes.  Bayesian trend filtering, in this way, lives in the intersection of $\ell_1$ penalized regression, namely the (generalized) lasso, and the Gaussian process prior literature.  From this vantage point, we found two distinct bodies of research that offer the framework for a proof of the convergence rate of Bayesian trend filtering.

Bayesian trend filtering is closely related to the work of \citet{Mammen:1997} on locally adaptive regression splines.  To show that locally adaptive regression splines are convergent at the minimax rate, \citet{Mammen:1997} use the metric entropy calculations of \citet{Van-de-Geer:1990,Mammen:1991} and also interpolating properties of splines developed by \citet{Boor:2001}.  Building on this work, \citet{Tibshirani:2014} proves a similar conclusion for trend filtering.  On the other hand, Bayesian trend filtering uses a hierarchical model common to Gaussian process regression.  Some in the Gaussian process literature have a similar goal in mind: prove rates of convergence of posterior distributions that are based on Gaussian process priors \citep{Ghosal:2007,Van-der-Vaart:2008,Van-Der-Vaart:2011}.  This work on Gaussian process priors, which contains Bayesian trend filtering, notes that at the heart of the metric entropy proofs relating to nonparametric regression, there lies interpolating properties of the function space of interest \citep{Ghosal:2007}.  Though we were not able to prove such interpolation properties about the space of piecewise polynomials, this connection between penalized regression techniques and the Gaussian process regression methods is quite interesting.

Bayesian trend filtering has good potential in application.  Based on our simulations, we find that Bayesian trend filtering has good estimation and strong frequentist properties, as compared to the original trend filtering and a popular cubic smoothing spline method.  These benefits come from both the $\ell_1$ penalty, which acts similar to the total variation penalty, and the ability to estimate well the penalty parameter $\lambda$.  We also found that the generalized double Pareto conditional prior provides a substantial increase in Bayesian trend filtering's accuracy and coverage probablities.  Overall, Bayesian trend filtering decreases estimation error compared to the other methods we tested.

Bayesian trend filtering's benefits though come at some cost.  Bayesian trend filtering relies on a matrix inversion within the full conditional $[f|\cdot]$ at every iteration.  This matrix inverse is the most significant computational cost of Bayesian trend filtering, and it puts the method's computational complexity to be $\mathcal{O}(n^3)$.  Though, it should be reemphasized that depsite the computational burden, more information is gained from a Bayesian trend filtering fit.  Some in the Gaussian process literature developed means to avoid such an inverse in special cases, for instance see \citet{Vehtari:2007,Liu:2014}, but these methods are not directly applicable to Bayesian trend filtering.

A simplistic strategy to reduce the computation time for Bayesian trend filtering uses the Gibbs sampler's fast convergence rate.  From Proposition~\ref{theorem:ergodic} and from our simulations, we know that the Gibbs sampler of Bayesian trend filtering converges very quickly.  We could reduce computational complexity by sampling from the full conditional for $f$ every $m$th iteration, while sampling all other full conditionals every iteration.  This would reduce the computational burden of the full conditional $[f|\cdot]$, but at the same time produce larger effective sample sizes for the other parameters of interest.  We refit Bayesian trend filtering using this idea with $m=2$.  Table~\ref{tab:speeds} displays the computation times of each method as fit to the real data sets discussed above.  Bayesian tren filtering's estimate of the underlying function at each input $x_i$ changed very little between the two fits, sampling every iteration ($m=1$) verse sampling every other iteration ($m=2$).  The mean relative difference of the posterior mean of function evaluations at each input $x_i$ are $0.0028$ and $0.0009$, for the temperature and sunspot data, respectively.  

\begin{table}[ht]
\centering
\begin{tabular}{lcccccc}
  data set & BTF(m=1/m=2) & TF(k=$5$/$10$) & CSM & CSM-AR($1$) & BTREE\\
  \hline
  temperature  & $7/4$ & 10/20 & 0.2 & 0.5 & 9 \\
  sunspots  & 98/51 & 46/94 & 7 & 13 & 19 \\
\hline
\end{tabular}
\caption{Computation times in seconds for each method against each real data set.}
\label{tab:speeds}
\end{table}

The proposed hierarchical model for Bayesian trend filtering is essentially the complete framework for Bayesian generalized lasso.  With this many other models could be carried over into the Bayesian framework and with a variety of variations on the penalty presented here.  For instance, the elastic net penalty might improve the accuracy of Bayesian trend filtering.  Futher, Bayesian trend filtering itself could be modified in a number of interesting ways.  For instance, an additive model, $\mathbb{E}y_i = \sum_{j=1}^J f_j(x_{ij})$, is desirable.  However, because of the large computational complexity of Bayesian trend filtering it seems necessary to first rid this method of its matrix inversion.  Possibly some approximate Bayesian sampling technique is suitable.  This, together with the work on the minimax convergence rate proof via metric entropy methods, are left for future work.

\appendix

\section{Majorization-Minimization Algorithm}
\label{app:mm}

We offer an approximation to the objective function (\ref{eq:tf}).  The approximation uses the majorization-minimization techniques developed by \citet{Hunter:2005}.  Convergence, up to numerical precision, is nearly guaranteed since equation~(\ref{eq:tf}) is convex.

\begin{definition}[Majorization]
  Let $\theta^{[m]}$ be the $m^{th}$ iteration in a search for the minimum value of a function $f(\theta)$.  A function $g(\theta | \theta^{[m]})$ is said to majorize the real-valued function $f(\theta)$ at the point $\theta^{[m]}$ if 

  \begin{align*}
    g(\theta|\theta^{[m]}) & \geq f(\theta), \quad \forall \theta, \text{ and } \\
    g(\theta^{[m]} | \theta^{[m]}) & = f(\theta^{[m]}).
  \end{align*}
\end{definition}

Minimization of the function of interest is established by repeated minimization of the majorizor and some stopping criterion is satisfied, or when a maximum number of iterations is reached..  The stopping criteria considered here is that of stability of the estimates, i.e.\ the algorithm stops when $||f^{[m+1]} - f^{[m]}||_{\infty} < \tau$, where $\tau := 10^{-5}$ was chosen.
  
For some $\epsilon>0$ and specified value $\hat{\lambda}_{CV}$, the following is a majorization of (\ref{eq:tf})

\begin{equation*}
\begin{aligned}
  g(f|f^{[0]}) = ||y - f||_2^2 + \hat{\lambda}_{CV} \left\{ \sum_{i=1}^{n-k-1}\right. (||\Dk &f^{[0]}||_1)_i  - \epsilon \log \left( 1+\frac{(||\Dk f^{[0]}||_1)_i}{\epsilon} \right)  \\ 
& \left. + \frac{ \{(||\Dk f||_1)_i - (||\Dk f^{[0]}||_1)_i\}^2 }{2 (||\Dk f^{[0]}||_1)_i + \epsilon} \right\}.
\end{aligned}  
\end{equation*}

\noindent Unfortunately, the $\epsilon$ is not easily avoided as division by zero is otherwise encouraged.  Numerical precision becomes more and more of an issue as smaller values of $\epsilon, \tau$ are chosen.  Despite these issues with this approximation strategy, in our experience the mean absolute difference between this approximate solution and \texttt{genlasso::trendfilter}'s exact calculation was generally around $10^-3$.

\bibliographystyle{plainnat}
\bibliography{refs}
\end{document}